\documentclass{ws-procs975x65}

\usepackage{graphicx,epsfig,latexsym,amsmath,feynmf}

\begin{document}

\title{Subleading Spin-Orbit Correction to the Newtonian Potential \\ in Effective Field Theory Formalism}

\author{Delphine L. Perrodin}

\address{Department of Physics and Astronomy, Franklin \& Marshall College,\\
Lancaster, PA 17604, USA \\
$^*$E-mail: delphine.perrodin@fandm.edu}

\begin{abstract}
We study the gravitational dynamics in the early inspiral phase of coalescing compact binaries using Non-Relativistic General Relativity (NRGR) - an effective field theory formalism based on the post-newtonian expansion, but which provides a consistent lagrangian framework and a systematic way in which to study binary dynamics and gravitational wave emission. We calculate in this framework the spin-orbit correction to the newtonian potential at 2.5 PN. 
\end{abstract}


\bodymatter

%\section*{}
\vspace{1cm}
First introduced by Goldberger and Rothstein \cite{rothstein}, Non-Relativistic General Relativity (NRGR) provides a consistent framework using a lagrangian formalism where observables can in principle be calculated at any order in the velocity expansion. Corrections to the newtonian potential, such as the spinless 1PN and 2PN terms  have been successfully calculated using the NRGR framework \cite{rothstein, Ross}, in perfect agreement with previous post-newtonian calculations. In addition, Porto has developed a formalism which takes into account the spins of each of the compact objects \cite{rafael}. Porto showed that one can allow the local frame basis to rotate by adding rotational degrees of freedom ($e_I^{\mu}$) to the worldline action. The generalized angular velocity is defined by $\Omega^{\mu\nu} = e^{\nu J} \frac{De_J^{\mu}}{d\lambda}$, and the spin tensor $S_{\mu\nu}$ is introduced as the conjugate momentum to $\Omega_{\mu\nu}$. The worldline action, which is fixed by reparametrization invariance, is of the form \cite{rafael}:
\begin{equation}
S = - \sum_i \left( \int{ p_i^{\mu} u_{\mu}^i d\lambda_i} + \int{\frac{1}{2} S_i^{\mu\nu} \Omega_{\mu\nu}^i d\lambda_i} \right).
\end{equation}
In order to maintain the correct number of degrees of freedom, a spin supplementary condition (SSC) is added to the equations of motion. Using this formalism, Porto \& Rothstein calculated spin(1)-spin(2) and spin-squared couplings at leading order (2PN) and next-to-leading order (3PN)  \cite{3PN, spin1spin1}, in agreement with the post-newtonian results from Schaefer et al. \cite{schaefer3PN, schaeferspin1spin1}. The coupling of compact object spin with the angular momentum of the binary ({\it spin-orbit} coupling) was calculated by Porto \& Rothstein at leading order (1.5PN) \cite{rafael}:
\begin{equation}
V_{1.5PN}^{so} = \frac{G \, m_2}{r^2} n^j \left( S_1^{j0} + S_1^{jk} (v_1^k-2 v_2^k) \right) + 1 \leftrightarrow 2
\end{equation}
The next-to-leading order spin-orbit term at 2.5PN had not yet been calculated in NRGR. Calculations of this term have been done using post-newtonian techniques by Buonanno et al. \cite{buonanno}, Tagoshi et al. \cite{tagoshi} and Damour et al. \cite{PNdamour}. We note that Kol \& Smolkin have also explored spin-orbit at 2.5PN using NRGR methods and the so-called Kol variables. However they have so far only presented diagrams \cite{smolkin}. In this paper, we use the same spin formalism and methods that Porto \& Rothstein used to calculate the 3PN spin(1)-spin(2) term \cite{3PN}. At the next-to-leading order in spin-orbit, we have linear terms for the spin (Eq. 3), as well as quadratic terms (2 gravitons coupling to the spin term) (Eq. 4):
\begin{eqnarray}
L & = & \frac{1}{2 M_p} \, H_{\alpha\gamma, \beta} \, S^{\alpha\beta} \, u^{\gamma} \\
L & = & \frac{1}{4 M_p^2} \, S^{\beta\gamma} \, u^{\mu} \, H_{\gamma}\,^{\lambda} \, ( \frac{1}{2} H_{\beta\lambda,\mu} + H_{\mu\lambda,\beta} - H_{\mu\beta,\lambda} )
\end{eqnarray}
At 2.5PN, the potential is represented by a combination of Feynman diagrams:
%\vspace{0.001cm}
\section{Potential from single-graviton exchange}
%\vspace{0.1cm}
\begin{eqnarray}
V_{\text{single}} & = & \frac{G \, m_2}{2 \, r^2} \Big( n^k \, S_1^{ik} \{ v_1^i \vec{v_2}^2+2 \, v_2^i(\vec{v_2}^2-2 \, \vec{v_1}.\vec{v_2}) \} \nonumber \\ & & +S_1^{0k} \{ 4 \, r \, a_2^k-2 (v_1^k -2 \, v_2^k) \, \vec{v_2}.\vec{n}-3 \, n^k \, \vec{v_2}^2+4 \, n^k \, \vec{v_1}.\vec{v_2} \} \Big) 
\end{eqnarray}
\begin{figure}
\begin{center}
\begin{fmffile}{linear}
\newenvironment{Z}
  {\begin{fmfgraph*}(72.5,32.5)
  \fmfleft{i1,i2}  
  \fmf{vanilla}{i1,v1,o1}
  \fmf{dashes,tension=0}{v2,v1}
  \fmf{vanilla}{i2,v2,o2}
  \fmfright{o1,o2}}  
  {\end{fmfgraph*}}
\begin{Z}
  \fmfv{decor.shape=circle,decor.filled=empty,decor.size=3thick,label=$v^2$,l.d=.07w}{v2}
  \fmfv{decor.shape=circle,decor.filled=full,decor.size=1.5thick,label=$v^3$,l.d=.05w}{v1}
\end{Z}
\begin{Z}  
  \fmfv{decor.shape=circle,decor.filled=empty,decor.size=3thick,label=$v^3$,l.d=.07w}{v2}
  \fmfv{decor.shape=circle,decor.filled=full,decor.size=1.5thick,label=$v^2$,l.d=.05w}{v1}
\end{Z}
\begin{Z}
  \fmfv{decor.shape=circle,decor.filled=empty,decor.size=3thick,label=$v^4$,l.d=.07w}{v2}
  \fmfv{decor.shape=circle,decor.filled=full,decor.size=1.5thick,label=$v^1$,l.d=.05w}{v1}
\end{Z}
\begin{Z}  
  \fmfv{decor.shape=circle,decor.filled=empty,decor.size=3thick,label=$v^5$,l.d=.07w}{v2}
  \fmfv{decor.shape=circle,decor.filled=full,decor.size=1.5thick,label=$v^0$,l.d=.05w}{v1}
\end{Z}  
\end{fmffile}
\end{center}
\caption{Diagrams with single-graviton exchange. The blobs represent spin insertions. The leading order spin vertex introduces a $o(v^2)$ correction.}
\end{figure}
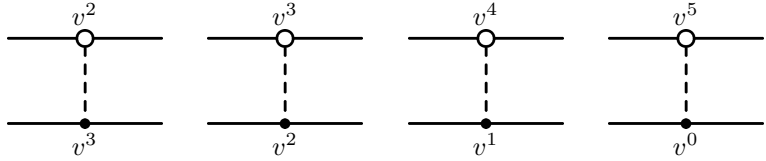
\section{Potential from propagator corrections}
%\vspace{0.1cm}
\begin{eqnarray}
V_{\text{prop}} & = & - \frac{G m_2}{2 \, r^2} \Big( S_1^{0k} \{ r \, a_2^k-r \, \vec{a_2}.\vec{n} \, n^k+ 2 \, v_2^k \, \vec{v_2}.\vec{n} + n^k (\vec{v_2}^2-3 \, (\vec{v_2}.\vec{n})^2) \} \nonumber \\ & &
+ S_1^{ik} \{ -2 \, r \, a_2^i (v_1^k-n^k \, \vec{v_1}.\vec{n}) - r \, a_1^i \, v_2^k + v_1^i \, \vec{v_1}.\vec{n} \, \vec{v_2}^k - 2 \, v_1^k \, v_2^i \, v_2^i \, \vec{v_2}.\vec{n} \nonumber \\ & & + n^k (r \, a_1^i \, \vec{v_2}.\vec{n}+(v_1^i-2 \, v_2^i) \, (\vec{v_1}.\vec{v_2}- 3 \, \vec{v_1}.\vec{n} \, \vec{v_2}.\vec{n})) \} \Big) 
\end{eqnarray}
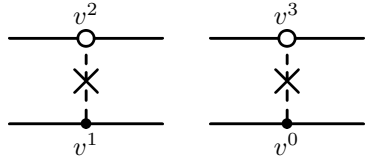
\begin{figure}[htp]
\begin{center}
\begin{fmffile}{prop}
\newenvironment{Z6}
  {\begin{fmfgraph*}(72.5,32.5)
  \fmfleft{i1,i2}  
  \fmf{vanilla}{i1,v1,o1}
  \fmf{dashes,tension=0.01}{v1,v3}
  \fmf{dashes,tension=0.01}{v3,v2}
  \fmf{vanilla}{i2,v2,o2}
  \fmfright{o1,o2}}  
  {\end{fmfgraph*}}
\begin{Z6}
  \fmfv{decor.shape=circle,decor.filled=empty,decor.size=3thick,label=$v^2$,l.d=.07w}{v2}
  \fmfv{decor.shape=circle,decor.filled=full,decor.size=1.5thick,label=$v^1$,l.d=.05w}{v1}
  \fmfv{decor.shape=cross,decor.filled=empty,decor.size=6thick}{v3}  
\end{Z6}
\begin{Z6}
  \fmfv{decor.shape=circle,decor.filled=empty,decor.size=3thick,label=$v^3$,l.d=.07w}{v2}
  \fmfv{decor.shape=circle,decor.filled=full,decor.size=1.5thick,label=$v^0$,l.d=.05w}{v1}
  \fmfv{decor.shape=cross,decor.filled=empty,decor.size=6thick}{v3}  
\end{Z6}
\end{fmffile}
\end{center}
\caption{Single-graviton diagrams with propagator corrections, which scale as $o(v^2)$. There are different ways to calculate this term.}
\end{figure}
\newpage
\section{Potential from quadratic diagrams}
%\vspace{0.05cm}
\begin{equation}
V_{\text{seagull}}  = - \frac{G^2 \, m_2}{r^3} \, n^k \, \Big( m_1 (S_1^{0k}+S_1^{ik} (v_1^i - 2 \, v_2^i))+ 2 \, m_2 S_1^{ik} (v_2^i-v_1^i) \Big)
\end{equation}
\begin{figure}[htp]
\begin{center}
\begin{fmffile}{seagull}
\newenvironment{Z2}
  {\begin{fmfgraph*}(58,26)
  \fmfleft{i1,i2}  
  \fmf{vanilla}{i1,v1,v3,o1}
  \fmf{dashes,tension=0}{v2,v1}
  \fmf{dashes,tension=0}{v2,v3}
  \fmf{vanilla}{i2,v2,o2}
  \fmfright{o1,o2}}  
  {\end{fmfgraph*}}
 \newenvironment{Z3}
  {\begin{fmfgraph*}(58,26)
  \fmfleft{i1,i2}  
  \fmf{vanilla}{i1,v2,o1}
  \fmf{dashes,tension=0}{v2,v1}
  \fmf{dashes,tension=0}{v2,v3}
  \fmf{vanilla}{i2,v1,v3,o2}
  \fmfright{o1,o2}}  
  {\end{fmfgraph*}}
\begin{Z2}
\fmfv{decor.shape=circle,decor.filled=empty,decor.size=3thick,label=$v^4$,l.d=.07w}{v2}
\fmfv{decor.shape=circle,decor.filled=full,decor.size=1.5thick,label=$v^1$,l.d=.05w}{v1}
\fmfv{decor.shape=circle,decor.filled=full,decor.size=1.5thick,label=$v^0$,l.d=.05w}{v3}  
\end{Z2}
\begin{Z2}
\fmfv{decor.shape=circle,decor.filled=empty,decor.size=3thick,label=$v^5$,l.d=.07w}{v2}
\fmfv{decor.shape=circle,decor.filled=full,decor.size=1.5thick,label=$v^0$,l.d=.05w}{v1}
\fmfv{decor.shape=circle,decor.filled=full,decor.size=1.5thick,label=$v^0$,l.d=.05w}{v3}  
\end{Z2}
%\\ [3\baselineskip]
\begin{Z3}
\fmfv{decor.shape=circle,decor.filled=empty,decor.size=3thick,label=$v^2$,l.d=.07w}{v1}
\fmfv{decor.shape=circle,decor.filled=full,decor.size=1.5thick,label=$v^2$,l.d=.05w}{v2}
\fmfv{decor.shape=circle,decor.filled=full,decor.size=1.5thick,label=$v^1$,l.d=.05w}{v3}  
\end{Z3}
\begin{Z3}
\fmfv{decor.shape=circle,decor.filled=empty,decor.size=3thick,label=$v^2$,l.d=.07w}{v1}
\fmfv{decor.shape=circle,decor.filled=full,decor.size=1.5thick,label=$v^3$,l.d=.05w}{v2}
\fmfv{decor.shape=circle,decor.filled=full,decor.size=1.5thick,label=$v^0$,l.d=.05w}{v3}  
\end{Z3}
\begin{Z3}
\fmfv{decor.shape=circle,decor.filled=empty,decor.size=3thick,label=$v^3$,l.d=.07w}{v1}
\fmfv{decor.shape=circle,decor.filled=full,decor.size=1.5thick,label=$v^2$,l.d=.05w}{v2}
\fmfv{decor.shape=circle,decor.filled=full,decor.size=1.5thick,label=$v^0$,l.d=.05w}{v3}  
\end{Z3}
\end{fmffile}
\end{center}
\caption{Quadratic "seagull" diagrams with double-graviton exchange. The leading order quadratic spin vertex scales as $o(v^4)$.}
\end{figure}
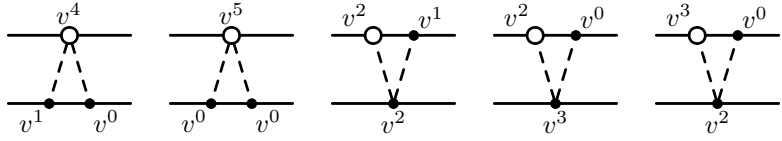
\section{Potential from 3-graviton diagrams}
%\vspace{0.05cm}
\begin{equation}
V_{\text{3-grav}} = \frac{G^2 \, m_2}{2 \, r^3} \, n^k \, \Big( 4 \, (m_1+m_2) S_1^{0k}-S_1^{ik} (3 \, m_2 v_1^i + (4 m_1+m_2) \, v_2^i) \Big)
\end{equation}
\begin{figure}[htp]
\begin{center}
\begin{fmffile}{three}
\newenvironment{Z4}
  {\begin{fmfgraph*}(72.5,32.5)
  \fmfleft{i1,i2}  
  \fmf{vanilla}{i1,v1,v3,o1}
  \fmf{dashes,tension=0.01}{v1,v4,v3}
  \fmf{dashes,tension=0.016}{v4,v2}
  \fmf{vanilla}{i2,v2,o2}
  \fmfright{o1,o2}}  
  {\end{fmfgraph*}}
\begin{Z4}
\fmfv{decor.shape=circle,decor.filled=empty,decor.size=3thick,label=$v^2$,l.d=.07w}{v2}
\fmfv{decor.shape=circle,decor.filled=full,decor.size=1.5thick,label=$v^1$,l.d=.05w}{v1}
\fmfv{decor.shape=circle,decor.filled=full,decor.size=1.5thick,label=$v^0$,l.d=.05w}{v3}
\fmfv{decor.shape=circle,decor.filled=full,decor.size=1thick}{v4}
\end{Z4}
\begin{Z4}
\fmfv{decor.shape=circle,decor.filled=empty,decor.size=3thick,label=$v^3$,l.d=.07w}{v2}
\fmfv{decor.shape=circle,decor.filled=full,decor.size=1.5thick,label=$v^0$,l.d=.05w}{v1}
\fmfv{decor.shape=circle,decor.filled=full,decor.size=1.5thick,label=$v^0$,l.d=.05w}{v3}
\fmfv{decor.shape=circle,decor.filled=full,decor.size=1thick}{v4}
\end{Z4}
\begin{Z4}
\fmfv{decor.shape=circle,decor.filled=empty,decor.size=3thick,label=$v^2$,l.d=.07w}{v2}
\fmfv{decor.shape=circle,decor.filled=full,decor.size=1.5thick,label=$v^0$,l.d=.05w}{v1}
\fmfv{decor.shape=circle,decor.filled=full,decor.size=1.5thick,label=$v^0$,l.d=.05w}{v3}
\fmfv{decor.shape=cross,decor.filled=empty,decor.size=6thick}{v4}
\end{Z4}
 \\[3\baselineskip]
\newenvironment{Z5}
  {\begin{fmfgraph*}(72.5,32.5)
  \fmfleft{i1,i2}  
  \fmf{vanilla}{i1,v2,o1}
  \fmf{dashes,tension=0.01}{v1,v4,v3}
  \fmf{dashes,tension=0.016}{v4,v2}
  \fmf{vanilla}{i2,v1,v3,o2}
  \fmfright{o1,o2}}  
  {\end{fmfgraph*}}
\begin{Z5}
\fmfv{decor.shape=circle,decor.filled=empty,decor.size=3thick,label=$v^2$,l.d=.07w}{v1}
\fmfv{decor.shape=circle,decor.filled=full,decor.size=1.5thick,label=$v^0$,l.d=.05w}{v2}
\fmfv{decor.shape=circle,decor.filled=full,decor.size=1.5thick,label=$v^1$,l.d=.05w}{v3}
\fmfv{decor.shape=circle,decor.filled=full,decor.size=1thick}{v4}
\end{Z5}
\begin{Z5}
\fmfv{decor.shape=circle,decor.filled=empty,decor.size=3thick,label=$v^2$,l.d=.07w}{v1}
\fmfv{decor.shape=circle,decor.filled=full,decor.size=1.5thick,label=$v^1$,l.d=.05w}{v2}
\fmfv{decor.shape=circle,decor.filled=full,decor.size=1.5thick,label=$v^0$,l.d=.05w}{v3}
\fmfv{decor.shape=circle,decor.filled=full,decor.size=1thick}{v4}
\end{Z5}
\begin{Z5}
\fmfv{decor.shape=circle,decor.filled=empty,decor.size=3thick,label=$v^3$,l.d=.07w}{v1}
\fmfv{decor.shape=circle,decor.filled=full,decor.size=1.5thick,label=$v^0$,l.d=.05w}{v2}
\fmfv{decor.shape=circle,decor.filled=full,decor.size=1.5thick,label=$v^0$,l.d=.05w}{v3}
\fmfv{decor.shape=circle,decor.filled=full,decor.size=1thick}{v4}
\end{Z5}
\begin{Z5}
\fmfv{decor.shape=circle,decor.filled=empty,decor.size=3thick,label=$v^2$,l.d=.07w}{v1}
\fmfv{decor.shape=circle,decor.filled=full,decor.size=1.5thick,label=$v^0$,l.d=.05w}{v2}
\fmfv{decor.shape=circle,decor.filled=full,decor.size=1.5thick,label=$v^0$,l.d=.05w}{v3}
\fmfv{decor.shape=cross,decor.filled=empty,decor.size=6thick}{v4}
\end{Z5}
\end{fmffile}
\end{center}
\caption{3-graviton diagrams. The leading order 3-graviton vertex scales as $o(v^2)$, while crosses correspond to $o(v)$ - corrections to the 3-graviton vertex. 3-graviton vertices were computed with the help of Rothstein's 3-graviton code \cite{code}.}
\end{figure}
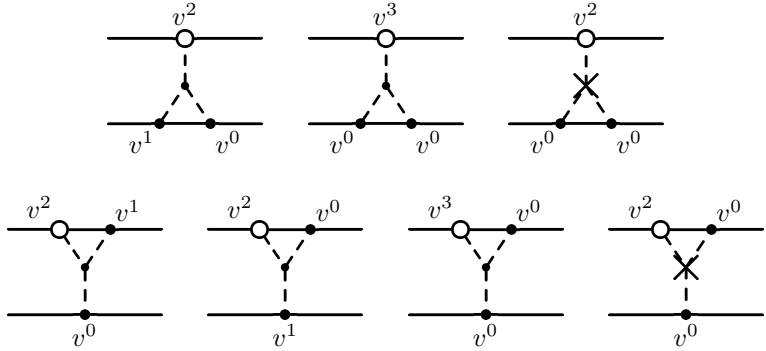
\section{Spin-orbit potential at 2.5PN}
From the sum of these diagrams, we extract the spin-orbit potential at 2.5PN:
\begin{eqnarray}
V_{2.5PN}^{so} & = & \frac{G \, m_2} {2 \, r^2} \Big( S_1^{0k} \, \{  3 \, r \, a_2^k + r \, \vec{a_2}.\vec{n} \, n^k +2 \, \vec{v_2}.\vec{n} \, (v_2^k - v_1^k) + n^k \, (3 \, (\vec{v_2}.\vec{n})^2 +4 \, (\vec{v_1}.\vec{v_2}-\vec{v_2}^2)) \}  \nonumber \\ & & + S_1^{ik} \{ 2 \, r \, a_2^i \, (v_1^k-n^k \, \vec{v_1}.\vec{n}) + r \, a_1^i \, v_2^k - v_1^i \, \vec{v_1}.\vec{n} \, v_2^k + \, 2 \, v_1^k \, v_2^i \, \vec{v_2}.\vec{n} \nonumber \\  &  & +n^k \, ((- r \, a_1^i + 3 \, v_1^i \, \vec{v_1}.\vec{n}-6 \, \vec{v_1}.\vec{n} \, v_2^i) \, \vec{v_2}.\vec{n} - (\vec{v_1}.\vec{v_2} -\vec{v_2}^2) \, (v_1^i+2
   \, v_2^i)) \} \nonumber \\ &  & + \frac{G}{r} n^k (2 \, m_1 (S_1^{0k}-S_1^{ik} \, v_1^i)+ m_2 (4 \, S_1^{0k}+S_1^{ik} (v_1^i-5 v_2^i))) \Big) + 1 \leftrightarrow 2 
\end{eqnarray}

\newpage
Using the methods of NRGR, we have calculated the next-to-leading order (2.5PN) spin-orbit correction to the gravitational potential in inspiralling compact binaries. Because of the different gauges, a direct comparison of this potential with that of Buonanno et al. \cite{buonanno} is not possible at this stage. The calculation of a gauge-invariant quantity such as the energy of circular orbits is the object of an upcoming publication. Other subtleties such as different definitions for spin also need to be considered when comparing results. Spin is here defined in the local frame of each compact object, as opposed to the PN frame used by Buonanno et al. \cite{buonanno}. We note that while the second version of this paper was being written, Porto computed the NRGR potential at the same order and cross-checked it against PN calculations using the spin precession equation. \cite{rafaelrecent}. 

\section*{Acknowledgements}
I would like to thank my PhD advisor Sean Fleming and Ira Rothstein for their support and useful discussions. I am also grateful to Rafael Porto and Andreas Ross for discussions and checks on Feynman diagram calculations. This work was supported in part by grant NSF AST 0748580.

%\verb|ws-procs975x65.bst| for references.
 
%\bibliographystyle{ws-procs95x65}
%\bibliography{ws-pro-sample}

\begin{thebibliography}{11}

\bibitem{rothstein} W. Goldberger and I. Rothstein, {\em Phys.\ Rev.}{\bf D73}, 104029 (2006).
\bibitem{Ross} J. Gilmore and A. Ross, {\em Phys.\ Rev.}{\bf D78}, 124021 (2008).
%\bibitem{kol} B. Kol and M. Smolkin, {\em Phys.\ Rev.}{\bf D77}, 064033 (2008).
\bibitem{rafael} R. Porto, {\em Phys.\ Rev.}{\bf D73}, 104031 (2006).
\bibitem{3PN} R. Porto and I. Rothstein, {\em Phys.\ Rev.}{\bf D78}, 044012 (2008).
\bibitem{spin1spin1} R. Porto and I. Rothstein, {\em Phys.\ Rev.}{\bf D78}, 044013 (2008).
\bibitem{schaefer3PN} J. Steinhoff, S. Hergt and G. Schaefer, {\em Phys.\ Rev.}{\bf D77}, 081501 (2008).
\bibitem{schaeferspin1spin1} J. Steinhoff, S. Hergt and G. Schaefer, {\em Phys.\ Rev.}{\bf D78}, 101503 (2008).
\bibitem{buonanno} G. Faye, L. Blanchet and A. Buonanno, {\em Phys\ Rev.}{\bf D74}, 104033 (2006).
\bibitem{tagoshi} H. Tagoshi, A. Ohashi and B. Owen, {\em Phys.\ Rev.}{\bf D63}, 044006 (2001).
\bibitem{PNdamour} T. Damour, P. Jaranowski and G. Schaefer, {\em Phys.\ Rev.}{\bf D77}, 064032 (2008); {\em Phys.\ Rev.}{\bf D78}, 024009 (2008).
\bibitem{smolkin} B. Kol and M. Smolkin, {\em Class.\ Quant.\ Grav.}{\bf 25}, 145011 (2008).
\bibitem{code} I. Rothstein, http://www-hep.phys.cmu.edu/workshop
\bibitem{rafaelrecent} R. Porto, arXiv:1005.5730 (gr-qc)

\end{thebibliography}
 
\end{document}